# STUDIES ON ACCESS: A REVIEW

Philip M. Davis
December, 2009

Research on the accessibility of the scientific literature follows two main methodological approaches: the first is based on surveying researchers on their recollections, perceptions and desires of the journal publishing system; the second is based on unobtrusive studies of what scientists read and cite. Both approaches have their strengths and weaknesses.

Surveys can gather the responses of thousands of individuals and allow a researcher to generalize the results over a target population. In-depth interviews, while limited in their generalizability, can explore a topic in more detail and draw out values and motivations from a respondent. Poorly constructed questionnaires, however, can mislead respondents and result in biased results. Similarly, interviewees may be prompted to provide what researchers want to hear or what scientists ought to believe, leading to significant response bias. For example, since one of the central values of science is openness (Merton, 1942), scientists may be supportive of the phrase 'open-access' in spite of the ambiguity in how the term is used. Moreover, researchers have different – and often competing – interests when responding as *authors* or as *readers* (Mabe & Amin, 2002). Authors want to publish more: readers want to read less. This poses a problem for understanding the needs of researchers and makes the context of the study immensely important.

Unobtrusive measures (such as counting article downloads or measuring citations) are a more direct approach to measuring what scientists *actually do* and not they say they do. These studies, however, are unable to answer questions such as *why* an article was downloaded or cited. Clearly, both types of studies are required to develop a more complete picture of the state of access to the scientific literature.

In reviewing the literature, there is surprising consistency in the conclusions of these studies: *access to the published literature is improving, and those who generate knowledge view access issues as largely unimportant*. We should emphasize the phrase "those who generate knowledge," since there has been very little work on the dissemination of scientific information to those who use – but do not contribute to – the literature (i.e. teachers, medical practitioners, industrial researchers, and the lay public).

Moreover, most studies have focused on access to the formal, published literature and assume that access is provided either directly from the publisher or through a library intermediary. We should not ignore the many informal ways academics share documents among informal networks of peers. Lastly, we should understand that most of the surveys and interviews cited below were conducted prior to the recent economic downturn, which have resulted in significant material reductions in major academic libraries.

*Survey Studies*

Since 1977, periodic surveys of the reading and information-seeking patterns of U.S.-based scientists have been performed allowing for longitudinal trends to be reported (D. W. King & Tenopir, 1999; D. W. King, Tenopir, Montgomery, & Aerni, 2003; Tenopir & King, 2000a, 2000b, 2002; Tenopir & King, 2008; Tenopir et al., 2003; Tenopir, King, Edwards, & Wu, 2009). Over the previous three decades, Tenopir reports, the average number of readings per scientist has been rising while the time spent finding and reading an article has been steadily decreasing. These studies have also indicated that scientists are reading from a broader group of journals and extending their readership into the older literature, a trend that Tenopir and King attribute to the digitization of the journal literature and the creation of electronic archives. Scientists in the United States are relying primarily on institutional (library) access to journal collections although do rely on informal sources (such as preprint serves or colleagues) for some of their literature needs. A recent survey of researchers in India illustrates the importance of informal sources of scientific literature in countries where institutional and library access is more restricted (Gaulé, 2009). In the previous three months, Gaulé describes, 84% of Indian researchers reported either contacting an author or a colleague for a copy of an article when formal routes of access were unavailable.

A large, international survey of senior authors of scientific papers in 2005 revealed much about the values of researchers (Rowlands & Nicholas, 2005). In deciding upon which journals to submit their work, factors such as reputation of the journal, readership, Impact Factor, and speed of publication ranked as the top concerns of authors. Conversely, permission to post a copy of the article or retaining copyright were ranked last. At the time of this study, there seemed to be little knowledge of what open access meant, some authors claiming to have published in open access journals when in fact they had not. Whereas the results of this survey reflected the views of over 5,000 authors, we should understand that the survey population consisted of a group of *corresponding authors* who were selected from the Institute for Scientific Information (ISI) author database. As a consequence, the results of this survey are biased toward senior authors who publish in higher impact journals. We should also be aware that the response rate of the survey was just over 7% and may reflect a more motivated, and thus opinionated, group of respondents.

A later report, focusing on a subset of researchers in immunology and microbiology (Rowlands & Olivieri, 2006), indicated that two-thirds (67%) of respondents indicated they either had ‘good’ or ‘excellent’ access to the literature, and that nearly 84% claimed that access is much better than it was 5 years ago. Nearly all (97%) of respondents reported that they were “very up-to-date with the current literature in their area.” Moreover, compared to other barriers, researchers did not rate problems in accessing the literature as significant impediments to their work. Based on a list of other considerations, access ranked 12 out of 16, just above a desire for more conferences and networking opportunities, better management and training and clearer ethical guidelines. Surveying a similar author population (and using the same access questions as Rowlands), Mark Ware reported that some 69% of respondents claimed having either 'good' or 'excellent' access to the literature, although this figure varied by region of the world (Ware, 2007, p. 8). The United States and Canada subgroup reported the highest satisfaction (85% 'good or excellent' access versus 3% 'poor'), with the 'Rest of the word' subgroup reporting significantly less satisfaction (53% versus 15% respectively). With the exception of the USA/Canada subgroup, the survey does not specify which countries make up each regional subgroup leading to difficult interpretation (e.g. Europe/M. East, Anglophone, Australasia versus Asia, and Rest of World).

In a recent survey of small and medium-sized enterprises in the United Kingdom, over 70% of respondents claimed that they had reasonably good access to the journal literature, with 60% further reporting that access was easier than it was 5 years ago (Ware, 2009). The study was based upon a convenience sample of businesses known to be users of the academic literature and reports a response rate of only 4%.

While these larger studies may be prone to several forms of bias, they appear to be confirmed by smaller, more rigorous studies of author preferences. Authors submitting manuscripts to the *British Medical Journal* reported that qualities such as Impact Factor, reputation, readership, speed of publication, and the quality of peer review played an important role in their decisions to submit a manuscript (Sara Schroter, Tite, & Smith, 2005), Schroter and Tite (2006). Consistent with Rowlands (2005, 2006), authors placed little if any priority on the access policy of the journal.

A series of in-depth interviews of faculty, librarians and administrators at the University of California, Berkeley revealed a disjoint between the views of librarians and faculty. "Unlike many faculty, librarians who were interviewed strongly perceive a crisis in scholarly communication" (C. J. King et al., 2006, p. 8). Faculty were, for the most part, focused on quality concerns in academic publishing and were insulated from the consideration of costs in the publication process. Disciplinary norms were the strongest determinant in how and where faculty published.

There is very little known on the effects of free access to the scientific literature on individuals outside of the research community. As mentioned above, all of the studies (with the exception of the UK study of small and medium-sized businesses) have focused on academic authors. To date, only one study on the clinical implications of access to the medical literature could be located (Hardisty & Haaga, 2008). In a pair of related experiments, researchers were interested in whether increased access would change the use of articles in clinical psychotherapy. Participating mental health professionals were provided with none of four access conditions: 1) a reference with no citation (the control); 2) a normal reference with citation; 3) a reference with an online linked citation; or 4) a reference with a linked citation to a free-access article. After one week, participants read a vignette on the same topic covered by the article and asked for recommendations for a medical intervention. In both studies, those participants in the free-access linked citation were more likely to report having read the article; however, in only one of the two studies did reading the article translate into making a recommendation consistent with the article. The researchers conclude that open access may increase the consumption of research articles, but that this may not necessarily influence clinical practice.

**Table A.** Summary of Key Access Studies

| Author | Survey type | Survey Population | Response rate | Key findings |
|---|---|---|---|---|
| Rowlands and Nicholas (2005) | Web-based survey | International sample of corresponding authors extracted by ISI author database. Survey conducted in 2005. | 5,513 of 76,790 invitations (7.2%) | In selecting a journal in which to publish, top concerns for authors were: Reputation of the journal, Readership, and Impact factor. Permission to post a copy of one's article and holding on to copyright were ranked last. |
| Rowlands and Olivieri (2006) | Web-based survey and interviews (phone, in-person) | Reanalysis of two prior author surveys undertaken in 2004 and 2005. Sample details not clear. | 3,695 sample and subsample of 92 immunologists and microbiologists | 67% of respondents (2004 survey) reported having either 'good' or 'excellent' access to the journal literature. 84% believe that access is improving. |
| King, C.J. et al (2006) | In-person interviews | 49 interviewees (31 faculty, 5 librarians, 2 campus-level administrators, 11 steering committee members). Faculty selected from 5 departments. 22/31 faculty are/were editors of scholarly journals. Interviews conducted 2005-6. | n/a | Disciplinary norms, the review and reward structure defined faculty views and behavior. Faculty were largely focused on quality issues in publishing (e.g. peer review), and were insulated from affordability issues. |
| Ware (2007) | Web-based survey | Recently published authors, reviewers, and editors of scientific journals. Data from ISI and journal websites. Survey conducted in 2007. | 3,040 of 39,232 (7.7%) | 69% reported having either 'good' or 'excellent' access to the journal literature; highest for USA and Canada (85%) and Australasia (84%), and lowest for rest of the world (53%) |
| Gaulé (2009) | Web-based survey | Corresponding India-based authors who had published in 2007 and extracted from ISI author database. Survey conducted in 2008. | 348 of 2,212 invitations (16%) | Reports high incidence of article requests from informal sources (peers, authors). Most article requests were honored. |
| Ware (2009) | Web-based survey | Subscription lists to trade magazines, corporate authors of STM articles, purchasers of individual journal articles (PPV). Survey conducted in 2009 | 1,131 of 26,390 invitations (4%) | For those who claimed that the research literature was important, 71% described their access as "fairly easy" or "very easy." 60% reported that access was easier than 5 years ago, 20% claimed it was worse. |

*Article Download and Citation Studies*

Article readership (as measured by publisher-reported fulltext downloads) is rising steadily and publisher journal packages have opened up access to huge numbers of journals that were previously inaccessible to college communities (Research Information Network, 2009). Publishers who offer these package deals view these data as an indication that they are providing increasing value to academic communities. Ease of access to a greater range of published literature is supported by surveys of scientists as mentioned above (Tenopir et al., 2009).

There is some dispute, however, on whether increased access has broadened the scope of cited material. Evans (2008), using a complex inferential model, suggests that commercial, online access to the literature is concentrating citations on a narrower group of more recent literature. Using a much simpler descriptive model, Larivière, Gingras, & Archambault (2008) report just the opposite.

Reporting on the first randomized controlled trial of open access publishing, Davis et al (2008) reported that freely-accessible articles received no more citations than subscription-access articles, although the freely-accessible treatment cohort did receive significantly more article downloads from a larger group of visitors. The lack of a citation differential implies that the traditional subscription model is efficient in disseminating published results to the research community and is consistent with the surveys of authors as reported above. The existence of a download advantage for freely-accessible articles may indicate a peripheral demand for scientific articles outside of the research community, although more research is required on illustrating who is accessing these articles and for what purpose.

If access to the published literature were a dire concern for researchers in developing countries, we would expect that open access journals would play a significant role in the citation behavior of researchers. A recent analysis of the citation patterns in 150 biology journals indicated that authors in developing countries are no more likely to write papers for open-access journals and are no more likely to cite open access articles (Frandsen, 2009). The Open Access coefficient in Frandsen's regression model was in fact negative (-4.51) although not significant (p=0.16). The small sample size in this study, however, permitted only large differences to be detectable. A much larger and robust comparative study between Swiss and Indian researchers revealed that articles written by Indian researchers had shorter reference lists and were more likely to cite articles from open access journals (Gaulé, 2009). The effect sizes reported by Gaulé, while statistically significant, were small. On average, reference lists were 6% shorter (less than 2 references) and contained 0.16 more citations to open access articles. Considering that Indian research institutions have far poorer access to the published literature than their Swiss counterparts, Indian researchers reported that they routinely requested copies of articles from authors and their peers at better-endowed institutions to supplement their literature needs. Some researchers admitted asking former students who moved to North American or European institutions for access to the literature.

A similar large-scale analysis of citation patterns by international authors revealed that free access to the published literature had a small but significant effect on citation behavior. Freely-accessible articles received about 8% more citations on average and twice that for poorer countries (Evans & Reimer, 2009). Commercial access to the literature, however, could explain a 40% increase in citations. It should be noted that Evans & Reimer were measuring the effect of *delayed free access* (when publishers make older articles freely available) and not the effect of self-archiving or author-pays open access publishing. A report released by Research4Life, an organization coordinating three programs (HINARI, AGORA and OARE) designed to provide free and highly-subsidized access to health, agricultural and environmental literature respectively to the poorest of the world's nations claimed that article production has increased in participating countries (Research4Life, 2009). The causal link between access and research output in this report is made on very rudimentary analysis without controlling for confounding variables (such as GDP or national expenditures on research and development) and that more rigorous analysis is needed before such a conclusion can be made.

In conclusion, the literature on the access indicates that access to the scientific literature is improving, and that compared to other research-related concerns, access is a low-priority concern. There is a dearth of research on whether free access to the scientific literature is making a difference in non-research contexts, such as in teaching, medical practice, industry and government policy making. Moreover, more work needs to be done on the dissemination of scientific papers through non-formal models such as peer-to-peer sharing networks.

**Table B.** Key papers on the citation effects of open access

| Author(s) | Study Design | Study Description | Main Results |
|---|---|---|---|
| (Lawrence, 2001) | Retrospective, Observational | 111,924 conference papers in computer sciences published between 1989-2000. Compared articles found freely on Internet with print-only access. Controls for venue. Online availability and citations from ResearchIndex. | Overall citation increase (mean=336%, median=158%). Greater citation effect reported for top 20 venues (mean=286%, median=284%) |
| (Schwarz & Kennicutt, 2004) | Retrospective, Observational | 1,679 papers published in the *Astrophysics Journal* in 1999 and 2002. 484 (61%) and 608 (72%) OA respectively. OA defined as any version of the article appearing in the astro-ph section of the arXiv. Citations counts from ISI. | Papers posted to astro-ph cited more than twice as often. Reports demographic differences among those who post articles to the arXiv compared to those who do not. |
| (Antelman, 2004) | Retrospective, Observational | 2,017 articles (802 OA (40%)) published in top 10 impact journals in philosophy, political science, engineering and mathematics between 1999-2002. OA defined as any version of article freely-available on Web. Compares mean citations across disciplines. Citation counts from ISI. | Mean OA citation differences (45-91%) depending on discipline. Citation differential more exaggerated for highly-cited articles |
| (Harnad & Brody, 2004) | Retrospective, Observational | 14 million articles published in physics between 1991 and 2001. OA defined as any version of article freely-available on the Web. Citation counts from ISI. Comparison methodology not defined. | Reports citation ratios between 2.5 and 5.8 in favor of OA. |
| (Metcalfe, 2005) | Retrospective, Observational | 7,089 articles (4,156 OA (59%)) published in 13 journals. OA defined as any copy of the article found in the astro-ph section of the arXiv. Citation counts from ISI. Basic comparison without controls. | Citation increases between 1.6 and 3.5 in favor of OA. As high as 5 for articles appearing in *Science* and *Nature*. |
| (Kurtz et al., 2005) | Retrospective, Observational | Articles published in 7 core astrophysics journals. OA defined as any copy found in the arXiv. Citation data from ADS system. Various analytic techniques employed. | Strong evidence that citation effect caused by self-selection and early-view effects. No evidence of of citation effect as a result of OA. |
| (Eysenbach, 2006) | Prospective, Observational | 1,492 articles (212 OA (14%)) published in *PNAS* in 2004. Author-pays OA articles freely-available from journal website for first 6-mo, after which all articles become freely-available. Controls for article and author characteristics in a logistic regression model. Citation counts from ISI. | OA articles were more likely to be cited than subscription-access articles between 0-6 mo, 4-10 mo, and 10-16 mo after publication (Odds ratios: 1.7, 2.1, 2.9 respectively) |
| (Davis & Fromerth, 2007) | Retrospective, Observational | 2,765 (511 OA (18.5%)) articles published in 4 math journals between 1997 and 2005. OA defined as any copy of the article present in the arXiv. Various analytic techniques with controls. Citation counts from MathSciNet. | OA articles received 35% more citations on average, more exaggerated for highly-cited articles. Self-selection argued as principle cause, not OA. |
| (Moed, 2007) | Retrospective, Observational | 18,757 articles published in 6 physics journals between 1992-2005 (1,913 OA (10.2%)). OA defined as any copy of the article found in the Condensed Matter section of the arXiv. Various analytic comparisons. Citation data from ISI. | No evidence of citation advantage as a result of access. Strong evidence that a quality differential between arXiv-deposited and non-deposited articles is responsible for citation effect. Evidence for earlier citation lifecycle for deposited articles |
| (Gaulé & Maystre, 2008) | Retrospective, Observational | 4,388 articles (17% OA) published in *PNAS* between 2004-2006. Author-pays OA articles freely-available from journal website for first 6-mo, after which all articles become freely-available. Linear regression model includes additional confounders over Eysenbach (2006) study. | When additional confounders (such as location of corresponding author and time of submission) were added to model, citation effect became insignificant. |
| (Davis et al., 2008) | Randomized controlled trial | 1,619 articles (247 OA (15%)) published in 11 physiology journals. Free access to articles from journal website. Controls for self-archiving. Logistic and negative-binomial regression analysis. Citation counts from ISI. | OA articles received more article downloads but were no more likely to be cited nor receive more citations within first year after publication |
| (Evans & Reimer, 2009) | Retrospective, Observational | 26 million articles published in 8,000 journals between 1998-2005. Measured effect of publisher-mediated free access with commercial online availability in Poisson regression model, controlling for journal volume effects. Citation counts from ISI. | Publisher-mediated free access increases citation rates by 8% on average (increasing for poorer countries), compared to 40% citation increase for commercial online access |
| (Norris, Oppenheim, & Rowland, 2008) | Retrospective, Observational | 4,633 articles (2,280 OA (49%)) published in ecology, applied math, sociology and economics. OA defined as any freely-available copy of article on Web. Simple comparisons with no controls. Citation data from ISI. | Average citation advantage ranged between 44%-88% depending upon field. |
| (Davis, 2009) | Retrospective, Observational | 11,013 articles (1,613 OA (15%)) published in 11 biomedical journals from 2003-2007. Author-pays OA articles freely-available from journal website with all journals offering delayed free access model. Linear regression models with article characteristics used as confounders. Citation counts from ISI. | Adjusted citation advantage of 17% for author-pays OA articles. Evidence of citation effect declining over time (from 32% in 2004 to 11% in 2007) |
| (Frandsen, 2009) | Retrospective, Observational | 150 journals in biology (34 of which were OA). Measures share of articles published by authors in developing countries and citations to OA journals. Linear regression. Citations from ISI. | Authors in developing countries are no more likely to publish their articles in OA journals and are no more likely to cite OA journals. Some evidence that OA journals tend to cite OA journals more frequently. |
| (Gaulé, 2009) | Retrospective, Observational | 43,150 articles in science and engineering published in 2007 by authors located in Switzerland and India. Linear regression with journal as fixed effect. | Indian reference lists were 6% shorter (2 fewer citations) and cite 50% more OA journals (0.16 more OA citations) than Swiss reference lists. Reference length differences were more pronounced in biology and medicine than in physics, engineering and chemistry. |
| (Lansingh & Carter, 2009) | Retrospective, Case-control study | 895 articles published in 6 journals in ophthalmology (3 OA, 3 subscription paired by Impact Factor) published in 2003. Multiple linear regression controlling for article characteristics. Citations from Scopus and Google Scholar. | Access status was not a significant predictor of citations when article characteristics were added to the regression model. |

**References**

Antelman, K. 2004. Do Open-Access Articles Have a Greater Research Impact? *College & Research Libraries* 65: 372-382, http://eprints.rclis.org/archive/00002309/

Davis, P. M. 2009. Author-choice open access publishing in the biological and medical literature: a citation analysis. *Journal of the American Society for Information Science and Technology* 60: 3-8, http://dx.doi.org/10.1002/asi.20965

Davis, P. M., & Fromerth, M. J. 2007. Does the arXiv lead to higher citations and reduced publisher downloads for mathematics articles? *Scientometrics* 71: 203-215, http://dx.doi.org/10.1007/s11192-007-1661-8

Davis, P. M., Lewenstein, B. V., Simon, D. H., Booth, J. G., & Connolly, M. J. L. 2008. Open access publishing, article downloads and citations: randomised trial. *BMJ* 337: a568, http://dx.doi.org/10.1136/bmj.a568

Evans, J. A. 2008. Electronic Publication and the Narrowing of Science and Scholarship. *Science* 321: 395-399, http://dx.doi.org/10.1126/science.1150473

Evans, J. A., & Reimer, J. 2009. Open Access and Global Participation in Science. *Science* 323: 1025-, http://dx.doi.org/10.1126/science.1154562

Eysenbach, G. 2006. Citation Advantage of Open Access Articles. *PLoS Biology* 4, http://dx.doi.org/10.1371/journal.pbio.0040157

Frandsen, T. F. 2009. Attracted to open access journals: a bibliometric author analysis in the field of biology. *Journal of Documentation* 65: 58-82, http://dx.doi.org/10.1108/00220410910926121

Gaulé, P. 2009. Access to scientific literature in India. *Journal of the American Society for Information Science and Technology* 12: 2548-2553, http://dx.doi.org/10.1002/asi.21195

Gaulé, P., & Maystre, N. (2008). *Getting cited: does open access help?* : CEMI Working Paper 2008-007, http://ilemt.epfl.ch/repec/pdf/cemi-workingpaper-2008-007.pdf

Hardisty, D. J., & Haaga, D. A. F. 2008. Diffusion of treatment research: does open access matter? *Journal of Clinical Psychology* 64: 821-839, http://dx.doi.org/10.1002/jclp.20492

Harnad, S., & Brody, T. 2004. Comparing the Impact of Open Access (OA) vs. Non-OA Articles in the Same Journals. *D-Lib Magazine* 10, http://www.dlib.org/dlib/june04/harnad/06harnad.html

King, C. J., Harley, D., Earl-Novell, S., Arter, J., Lawrence, S., & Perciali, I. (2006). *Scholarly Communication: Academic Values and Sustainable Models*. Berkeley, CA: Center for Studies in Higher Education, University of California, Berkeley,

http://cshe.berkeley.edu/publications/docs/scholarlycomm_report.pdf

King, D. W., & Tenopir, C. 1999. Using and Reading Scholarly Literature. *Annual Review of Information Science and Technology* 34: 423-477,

King, D. W., Tenopir, C., Montgomery, C. H., & Aerni, S. E. 2003. Patters of Journal Use by Faculty at Three Diverse Universities. *D-Lib Magazine* 9, http://www.dlib.org/dlib/october03/king/10king.html

Kurtz, M. J., Eichhorn, G., Accomazzi, A., Grant, C., Demleitner, M., Henneken, E., et al. 2005. The effect of use and access on citations. *Information Processing and Management* 41: 1395-1402, http://arxiv.org/abs/cs/0503029v1

Lansingh, V. C., & Carter, M. J. 2009. Does Open Access in Ophthalmology Affect How Articles are Subsequently Cited in Research? *Ophthalmology* 116: 1425-1431, http://dx.doi.org/10.1016/j.ophtha.2008.12.052

Larivière, V., Gingras, Y., & Archambault, É. 2008. The decline in the concentration of citations, 1900-2007. *Journal of the American Society for Information Science and Technology* 60: 858-862, http://dx.doi.org/10.1002/asi.21011

Lawrence, S. 2001. Free online availability substantially increases a paper's impact. *Nature* 411: 521, http://dx.doi.org/10.1038/35079151

Mabe, M. A., & Amin, M. 2002. Dr. Jekyll and Dr. Hyde: author-reader asymmetries in scholarly publishing. *ASLIB Proceedings* 54: 149-157, http://dx.doi.org/10.1108/00012530210441692

Merton, R. K. 1942. Science and Technology in a Democratic Order. *Journal of Legal and Political Sociology* 1: 115-126,

Metcalfe, T. S. 2005. The rise and citation impact of astro-ph in major journals. *Bulletin of the American Astronomical Society* 37: 555-557, http://arxiv.org/abs/astro-ph/0503519v1

Moed, H. F. 2007. The effect of 'Open Access' upon citation impact: An analysis of ArXiv's Condensed Matter Section. *Journal of the American Society for Information Science and Technology* 58: 2047-2054, http://dx.doi.org/10.1002/asi.20663

Norris, M., Oppenheim, C., & Rowland, F. 2008. The citation advantage of open-access articles. *Journal of the American Society for Information Science and Technology* 59: 1963-1972, http://dx.doi.org/10.1002/asi.20898

Research4Life. (2009). *Research Output in Developing Countries Reveals 194% Increase in Five Years*, http://www.who.int/hinari/Increase_in_developing_country_research_output.pdf

Research Information Network. (2009). *E-journals: their use, value and impact*, http://www.rin.ac.uk/files/E-journals_use_value_impact_Report_April2009.pdf

Rowlands, I., & Nicholas, D. (2005). *New journal publishing models: an international survey of senior researchers*: CIBER, http://www.ucl.ac.uk/ciber/ciber_2005_survey_final.pdf

Rowlands, I., & Olivieri, R. (2006). *Journals and Scientific Productivity: A case study in immunology and microbiology*. London: Publishing Research Consortium, http://www.publishingresearch.net/documents/Rowland_Olivieri_summary_paper.pdf

Schroter, S., & Tite, L. 2006. Open access publishing and author-pays business models: a survey of authors' knowledge and perceptions. *Journal of the Royal Society of Medicine* 99: 141-148, http://dx.doi/org/10.1258/jrsm.99.3.141

Schroter, S., Tite, L., & Smith, R. 2005. Perceptions of open access publishing: interviews with journal authors. *BMJ* 330: 756-, http://dx.doi.org/10.1136/bmj.38359.695220.82

Schwarz, G. J., & Kennicutt, R. C. J. 2004. Demographic and Citation Trends in Astrophysical Journal papers and Preprints. *Bulletin of the American Astronomical Society* 36: 1654-1663, http://arxiv.org/abs/astro-ph/0411275v1

Tenopir, C., & King, D. (2000a). Information-Seeking and Readership Patterns. Chapter 8. In *Toward Electronic Journals: Realities for Scientists, Librarians, and Publishers* (pp. 488). Washington: Special Libraries Association.

Tenopir, C., & King, D. (2000b). Readership of Scientific Scholarly Journals. Chapter 7. In *Toward Electronic Journals: Realities for Scientists, Librarians, and Publishers.* (pp. 488). Washington: Special Libraries Association.

Tenopir, C., & King, D. 2002. Reading behaviour and electronic journals. *Learned Publishing* 15: 259-265, http://dx.doi.org/10.1087/095315102760319215

Tenopir, C., & King, D. W. 2008. Electronic Journals and Changes in Scholarly Article Seeking and Reading Patterns. *D-Lib Magazine* 14, http://www.dlib.org/dlib/november08/tenopir/11tenopir.html

Tenopir, C., King, D. W., Boyce, P., Grayson, M., Zhang, Y., & Ebuen, M. 2003. Patterns of Journal Use by Scientists Through Three Evolutionary Phases. *D-Lib Magazine* 9, http://www.dlib.org/dlib/may03/king/05king.html

Tenopir, C., King, D. W., Edwards, S., & Wu, L. 2009. Electronic journals and changes in scholarly article seeking and reading patterns. *ASLIB Proceedings* 61: 5-32, http://dx.doi.org/10.1108/00012530910932267

Ware, M. (2007). *Peer review in scholarly journals: Perspectives of the scholarly community -- an international study*. Bristol, UK: Publishing Research Consortium,

http://www.publishingresearch.net/documents/PeerReviewFullPRCReport-final.pdf

Ware, M. (2009). *Access by UK small and medium-sized enterprises to professional and academic information*. Bristol, UK: Publishing Research Consortium, http://www.publishingresearch.net/documents/SMEAccessResearchReport.pdf